\documentclass[twocolumn,showpacs,preprintnumbers,amsmath,amssymb,english]{revtex4}

   \usepackage{graphicx}
\usepackage{dcolumn}
\usepackage{bm}
\begin{document}

\title
{Measurement of binding energy of negatively charged excitons in GaAs/Al$_{0.3}$Ga$_{0.7}$As  quantum wells\\}

\author{V.V. Solovyev} 
 \email{vicsol@issp.ac.ru} 
\author{I.V. Kukushkin \\}

\affiliation{
\it Institute of Solid State Physics, Russian Academy of Sciences, Chernogolovka, 142432 Russia \\
\it and Max-Planck-Institut fur Festkorperforschung, Heisenbergstrasse 1, 70569 Stuttgart, Germany}
\date{\today}

\begin{abstract}

We report a photoluminescence study of electron-hole complexes in specially designed semiconductor heterostructures. Placing a remote dilute layer of donors at different distances \itshape d \normalfont from the quantum well leads to the transformation of luminescence spectra of neutral ($X$) and negatively charged ($X^{-}$) excitons. The onset of an additional spectral line and its energy dependence on \itshape d \normalfont allows us to unambiguously relate the so-called $X^{-}$ trion state with charged excitons bound on charged donors in a barrier. The results indicate the overestimation in free-trion binding energies from previous studies of GaAs/Al$_{0.3}$Ga$_{0.7}$As quantum wells, and give their corrected values for QWs of width $200$ and $300$ \AA \space in the limiting case of infinitely distant donors.

\end{abstract}

\maketitle
\smallskip

\medskip

The semiconductor counterparts of hydrogen atoms, excitons, were predicted to exist some 70 years ago by Wannier and Mott~\cite{bib:Wanniermott} and proved to be a real entity only two decades later~\cite{Gross}.  A further analogy with such hydrogen formations as $H^{-}$ and $H_{2}^{+}$ has a similar but even more lingering history. Foreseen in 1958 by Lampert~\cite{Lampert}, trions have not been credibly identified in bulk materials ~\cite{BulkTrions}, mainly because of very small binding energies of three particles in 3D semiconductors. However, advances in growing artificial heterostructures of high purity gave hope of revealing these long-searched particles. In fact, the reduced dimensionality results in a noticeable consolidation of many-body complexes clamped together by increased Coulomb interaction~\cite{Stebe}. So trions should become more stable and thus experimentally observable.

Since the very first claims of observing 2D trions~\cite{Kheng} and demonstration of their several-particle nature~\cite{Finkelstein} there has been some dissension about the degree of their localization ~\cite{Bar-Joseph}. On the one hand, most experimental results were interpreted in terms of completely free particles. On the other hand, it has been reasonably surmised~\cite{Volkov1} that charged and several times as heavy as the conduction band electron, $X^{-}$ and a fortiori $X^{+}$ must be inevitably localized by charged residual impurities.  This argument seems even more convincing after taking into account the extremely low densities of light-created carriers, and the existent mobility edge of about $10^{9}$ cm$^{-2}$ even for agile electrons in the best available structures. As experimentally extracted parameters of trions (e.g. binding energy) can be seriously affected by disorder effects, the estimation of the strength of their localization is very important.

One attempt was made to study the lateral transport of dilute trion gas in an electric field ~\cite{Shields}. It has been shown that the $X^{-}$ emission spot experiences an in-plane shift under the applied bias. Though this phenomenon was claimed to be the most unambiguous proof of the free nature of trions, we note that it does not contradict the situation with stripping the charged particles off the localizing centers under a drift force.

Alternatively, one can probe the degree of localization of the charged trions by studying the temperature properties of their recombination line in PL-spectra, especially its shape~\cite{Volkov1}. The driving idea behind those measurements is based on the recoil mechanism. The recombination of direct-band neutral excitons with large total momentum $k$ (caused, for example, by their thermal motion) is forbidden by the momentum conservation law. The existence of a second electron in a free negatively charged exciton removes this restriction, since the total momentum of the complex can be transferred to the single electron remaining after recombination. As a result the energy of the emitted photon decreases by $E=E(k)[\frac{M}{m_e}-1]$, where $E(k)$ is the trion kinetic energy, $M$ is the trion mass, and $m_e$ is the electron mass. Substituting $M=2m_e + m_h$ , where $m_h$ is the hole mass, we obtain  $E=E(k)[\frac{m_h}{m_e}+1]$. Since $\frac{m_h}{m_e}>3$~\cite{Hannanov} we obtain  $E>4E(k)$. Thus the allowed range of recombination energy of three-particle complexes turns out to be greatly enhanced compared with the one of neutral excitons. Taking into account the exponential decrease of the oscillator strength of the $X^{-}$ recombination for higher k-vector \cite{OscStregths}, we come to the conclusion that the recombination line of a free trion should be asymmetrically broadened on the low-energy side by the amount 0.6 meV at 10 K. This result contradicts the experimental findings. In reality, the trion line has a symmetric shape and its width is close to 0.4 meV, which were found to be independent of temperature in the range 1.5-10 K. This discrepancy indicates that trion line corresponds to the radiative recombination of strongly localized excitonic complexes. In that case, momentum conservation during the recombination process is basically lifted, and this is the reason for the luminescence linewidth remains small and insensitive to the temperature.

\begin{figure}
	\centering
		\includegraphics{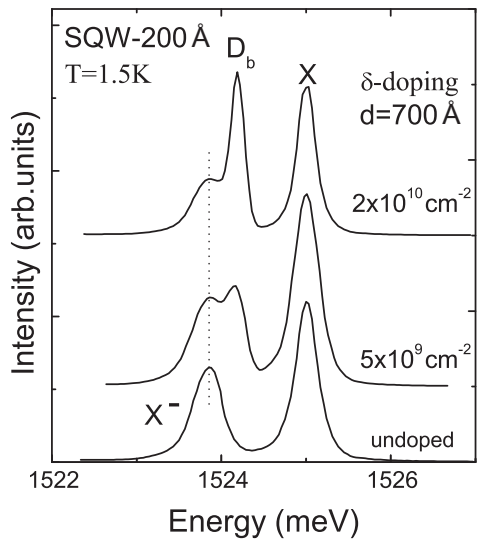}
		\caption{Effect of placing a distant dilute layer of donors on the exciton-trion photoluminescence from QW 200 \AA \space wide, for three doping levels. Note the appearance of an additional $D_{b}$-line in PL spectra.}
	\label{fig:Fig_1_improved}
\end{figure}

\begin{figure*}
	\centering
		\includegraphics[width=1.00\textwidth]{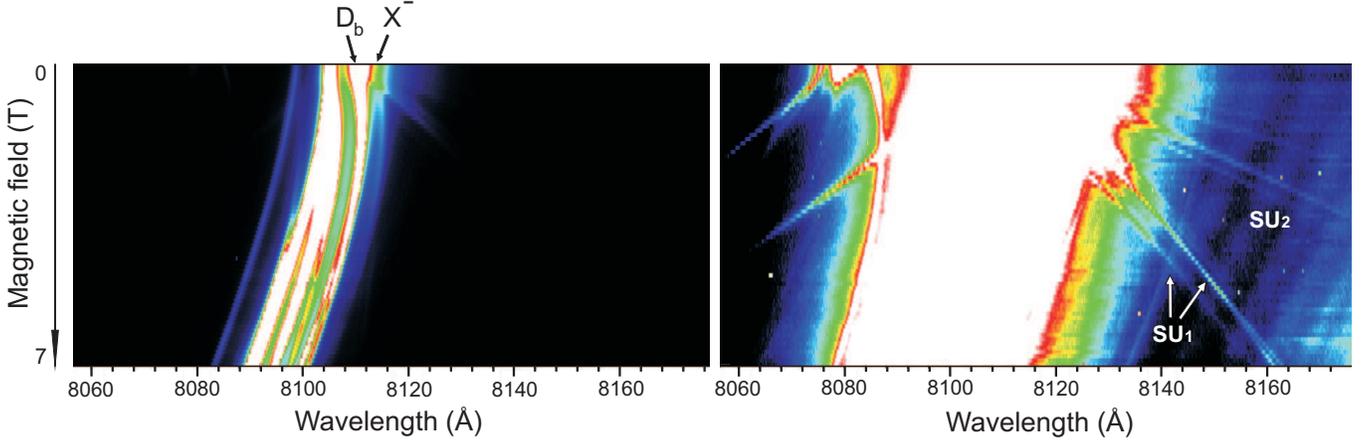}
	\caption{Images of magnetophotoluminescence spectra from a QW 300 \AA \space thick, with a doping layer at a distance 300 \AA, in a magnetic field varying from 0 to 7 Tesla. Shown are scaled to different intensity ranges representations of the same data. The left image demonstrates the most prominent features such as $X$, $X^{-}$ and $D_{b}$-lines, while on the right image one can see indications of shake-up satellites for both lines under discussion. Shake-up families SU1 and SU2 correspond to transfer to the remaining neutral donor of one and two cyclotron quanta, respectively.}
	\label{fig:Fig_2_improved}
\end{figure*}

In addition to recent experimental results suggesting the bound exciton picture of the trion state~\cite{Volkov2} we report here the direct evidence of $X^{-}$ localization on charged donors in a barrier. From our measurements we are able to determine the binding energy of free negatively charged excitons, and the results show that previous experimental values were greatly affected by disorder effects.

Since there is speculation about the influence of an unpredictable residual disorder on the properties of excitons, let us examine the effect of controllably introduced centers of localization. For this purpose we studied two series of nominally undoped and slightly doped GaAs/AlGaAs quantum wells (QW) $200$ and $300$ \AA \space thick. In doped QWs the remote $\delta$-doping by Si atoms was provided at distances $20$-$700$ \AA \space from the QW. All measurements were carried out at 1.5 K in a variable-temperature helium cryostat with a superconducting solenoid. Radiation of Ti-Sp laser excited photoluminescence through quartz fiber, the emitted light being collected in the same way and dispersed by a spectrometer equipped with a CCD-camera.

\begin{figure}
	\centering
		\includegraphics{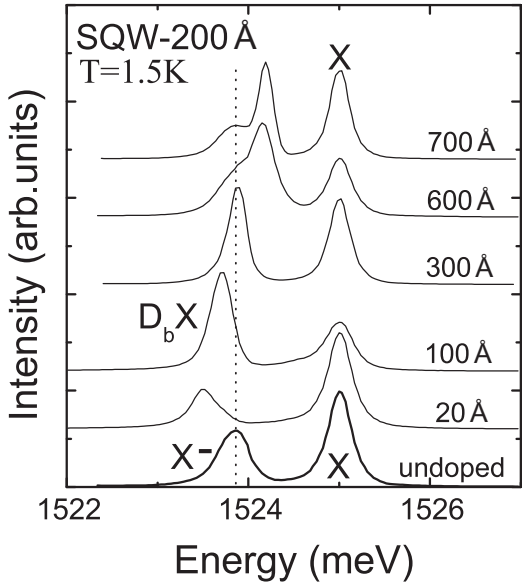}
		\caption{PL spectra for a set of samples that are nominally identical (single QW 200 \AA) but have different spacers \itshape d \normalfont to dopants. For closely spaced donors the $D_{b}$-line is red-shifted compared to the $X^{-}$-line, and moves to higher energies with increasing \itshape d \normalfont. The concentrations of intentional Si-doping are (from the top spectrum with \itshape d \normalfont=700 \AA \space to the bottom one with \itshape d \normalfont=20 \AA): 2x$10^{10}$, 1x$10^{10}$, 5x$10^{9}$, 2x$10^{9}$, 1x$10^{9}$ cm$^{-2}$. The residual 2D electron density is (1-3)x$10^{9}$ cm$^{-2}$ for all the samples.}
	\label{fig:Fig_3_improved}
\end{figure}

Experimentally, the growth of heterostructures that contain negligible residual doping is still far beyond of the capacity of the current MBE-technology. The state-of-art purity of the barrier material is about $10^{15}$ residual atoms per cm$^3$, and this leads to an accumulation of excessive charges on the heterointerface. The typical barrier thickness is about 200-400 nm, and this value sets the characteristic scale of 2D charge density (in dark) to be of order $10^{10}$ cm$^{-2}$. That is the point that should be kept in mind when discussing any properties of nominally undoped structures. Of course, it is possible to optically tune the total charge sign in the QW via a photodepletion effect, but without such measures the undoped QW cannot be considered having no excessive charges. Therefore typical luminescence spectra even from nominally undoped QWs almost always contain two intense spectral lines attributed to recombination of free excitons ($X$) and so-called trions ($X^{-}$ or $X^{+}$ depending on the charge of excessive carriers~\cite{spectra}; in the following we explain how to discriminate between these charge states by their magnetic field behavior). The spectral separation between these lines is usually used for determination of the trion binding energy~\cite{Bar-Joseph_review}. It is this value that would be greatly affected if additional binding on charged impurities did occur.

If we put a dilute layer of Si-donors at some distance from the QW, they become partly ionized due to transfer of electrons into the QW. So one may expect the possibility of negatively charged excitons localizing on these centers, with this process manifesting in the optical spectra. Fig.1 clearly indicates the onset of an additional spectral line due to the presence of remote doping at a distance of $700$ \AA \space from the $200$ \AA \space QW. This line, denoted as $D_{b}$, lies between the $X$ and $X^{-}$-lines, has a smaller linewidth compared with $X^{-}$, and its intensity grows monotonically as the doping level increases. These marked properties, along with some additional observations, give us conclusive evidence that $X^{-}$-state in fact corresponds to the strongly bound state of the trion and oppositely charged impurity.

The monotonical dependence of intensity on doping level clearly confirms that $D_{b}$-line is related to intentionally introduced doping. However its spectral position seems to be very contradictory if we assume that the $X^{-}$-line corresponds to the recombination of free-in-motion trions. In no way can free $X^{-}$ be more energetically favourable than its bound state $D_{b}$ The only explanation of these data is the conclusion that the optically detected $X^{-}$-state results in fact from the recombination of charged excitons bound on oppositely charged impurities. Or, virtually, from excitons bound on neutral donors. These residual centers of localization dwell in a barrier and bind excitons stronger than intentional doping does, due their greater proximity to the QW. A smaller linewidth of $D_{b}$ is then easily accounted for by less dispersive spatial distribution of donors in a remote Si layer.

Another demonstration of a common nature of $X^{-}$ and $D_{b}$ states is given by their behavior in magnetic field. It is well known that applying a magnetic field perpendicular to the QW plane leads to an appearance of a low-energy satellite for the $X^{-}$-line in the luminescence spectra ~\cite{Finkelstein}. The mechanism for this additional recombination (called `shake-up process') is connected with the ability of many-particle electron-hole complexes to recombine, leaving the remaining particle in different final states. If we consider a negatively charged exciton bound on ionized donor, then the optical transition in such a complex produces a photon and a neutral donor, the latter being left in either a ground or an excited state. The restriction of energy conservation therefore makes the emitted photon a sensitive probe for the actual condition of the remaining particle. Turning on a perpendicular magnetic field of rather high strength ~\cite{Field_note} reorganizes energy levels of the neutral donor into a discrete ladder of Landau levels (LL) for the neutralizing electron (with small corrections from an attractive potential of positive core). Hence this electron has an opportunity either to stay on the lowest LL or to jump one or more levels up. In optical spectra this process is revealed as a feeble line with energy shifted downward from the $X^{-}$-line by an integer number of cyclotron quanta ~\cite{nonobservance}. It is this experimental observation that was fundamental for proving the many-particle nature of the $X^{-}$, and, later, $X^{+}$-states. The slope of the shake-up energy shift gives cyclotron energy of the corresponding particle (electron or hole), thus unambiguously identifying it (and so the type of background charge) in experiment. 

 We made similar measurements on our samples in magnetic fields 0-7 Tesla. Image of Fig.2 clearly illustrates the presence of identical electron shake-up satellites for both $X^{-}$ and $D_{b}$-lines thus confirming their being the same.

Further evidence for the represented picture stems from the dependence of the $D_{b}$-line position on the distance between doping and QW. The transformation of luminescence spectra with varying \itshape d \normalfont is shown in Fig.3.

 For a very close doping layer (\itshape d \normalfont=$20$ \AA), the stronger binding of charged excitons compared to the undoped QW results in a red-shift of the $D_{b}$-line from the $X^{-}$-line (two lower spectra). With increasing d, we observe a gradual movement of the line of interest into higher energies. This fact demonstrates the presence of a significantly quenched but still appreciable attraction even from far-away ionized donors.

Nevertheless we are able to obtain an estimation for such a crucial property of a truly free trion as its binding energy, and very helpful in this procedure are the studied \itshape d \normalfont-dependencies of optical spectra. In fact, a bound trion becomes non-localized as its center of localization goes to infinity. Fig. 4 shows how the "`binding energy"' of the $X^{-}$-complex (which is just the energy separation between the $X$ and $X^{-}$-lines) changes with an increase in distance L, for two QW widths. Here L is the distance from the dopants to the QW center (L equals \itshape d \normalfont plus the QW halfwidth). Enhanced confinement in a narrower well promotes Coulomb clamping of electron-hole complexes, thus leading to stronger binding compared to the $300$ \AA \space well. From the monotonical dependence we find an estimate for the binding energy of the pure $X^{-}$ -state: the linear extrapolation to 1/L=0 gives values of approximately 0.7 meV and 0.5 meV for QW of width $200$ and $300$ \AA, respectively. These energies are two times smaller compared to previously reported data for GaAs/AlGaAs quantum wells ~\cite{Bar-Joseph, Bar-Joseph_review}, where the $X^{-}$-line was attributed to a pure, nonlocalized trion state.

\begin{figure}
	\centering
		\includegraphics{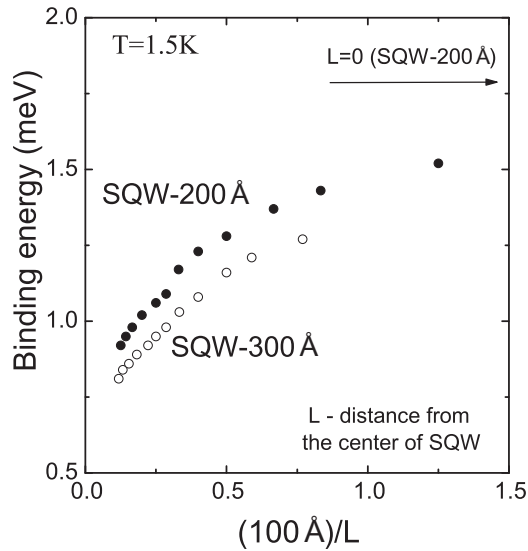}
		\caption{Binding energies of a negatively charged exciton on a remote ionized donor, extracted from PL studies of QWs with different distances to the doping layer, for two QW widths. Here L is the distance from dopants to the center of QW. The arrow indicates the upper limit for this value, which is just the case when the impurity is imbedded directly into the center of 200 \AA \space QW (L=0).}
	\label{fig:Fig_4_improved}
\end{figure}

It is appropriate to mention here disagreements between different theoretical predictions of trion binding energies. Early works ~\cite{theory1} used the simplest model that did not include details of band structure and electron-hole exchange interaction, and it yielded values of 2 and 1.8 meV for GaAs/AlGaAs QW $220$ and $300$ \AA \space wide, respectively. A more sophisticated approach ~\cite{theory2} dealt with stochastic variational method recruited to fully include the Coulomb interaction among the particles. As a result, binding energies of about 0.9-1 meV were obtained for the discussed QW widths. Reference ~\cite{theory5} presents similar findings. Finally, variational calculations within the configuration interaction method ~\cite{theory4} go down to values of 0.6-0.7 meV, when taking into account more than one electron QW solution. These latest theoretical predictions are perfectly matched with our results, thus indicating an intrinsically complex interplay of Coulomb interactions and a single-particle confinement potential of real quasi-2D systems.

In conclusion, we have investigated localization properties of charged excitons in GaAs/AlGaAs quantum wells. It is shown that even far-away ionized donors can bind charged 2D three-particle complexes consisting of two electrons and one hole. From the dependencies on the distance between the QW and doping atoms we extract values of binding energies for truly free trions.

Support from the RFBR is greatly acknowledged. The authors thank K. Eberl and M. Hauser for providing the high-quality samples.


\begin{thebibliography}{2002}
\bibitem{Field_note}
As the binding energy of the second electron in the $X^{-}$-complex is of order 1 meV, such a reorganization occurs in magnetic fields stronger than 1 T (with corresponding cyclotron energy is about 1.7 meV).
\end{thebibliography}
\end{document}